\documentclass{webofc} 
\usepackage[varg]{txfonts}   % Web of Conferences font
\usepackage{hyperref}
\usepackage{url}
\usepackage{lineno}
\usepackage{xcolor}
\hypersetup{colorlinks=true,citecolor=blue,urlcolor=blue,linkcolor=blue}

\begin{document}
\vspace{-5.0cm}

\title{Recent studies on heavy-flavor femtoscopy in heavy-ion collisions by STAR}

\author{\firstname{Priyanka} \lastname{Roy Chowdhury}\inst{1}\fnsep\thanks{\email{priyanka.roy_chowdhury.dokt@pw.edu.pl}}
(for the STAR Collaboration)
}

\institute{Warsaw University of Technology, Faculty of Physics, Koszykowa 75, 00-662 Warsaw, Poland }
 
\abstract{ At the initial stage of nuclear-nuclear collisions, heavy quarks are generated in hard partonic scatterings. This allows them to participate in the entire evolution of the heavy-ion collisions. During hadronization, different type of hadrons are produced including D mesons and light-flavoured hadrons, like pion ($\pi$), kaon ($K$), proton ($p$)  etc. We can observe different interactions between these hadrons based on the size of collision systems. 
Such as, hadron re-scattering, suppression of charm quarks and collective effects are weak or missing in $p+p$ system in comparison to $Au+Au$ or $Pb+Pb$ collision system~\cite{syst_size}.
 Femtoscopy is one of the most significant and unique tools for examining the final state interaction behaviors between correlated pair of particles at low momentum in a pair rest frame. It is also possible to explore the size and geometry of emission source through the measurements of femtoscopic correlation functions. 
Here we report the studies of correlations between neutral charmed meson ($D^0$ / $\overline{D^0}$) and identified charged hadrons ($\pi^{\pm}$, $K^{\pm}$, $p^{\pm}$) in Au+Au collisions at STAR experiment using femtoscopy technique. 
This is the first measurement of heavy-flavor femtoscopy in heavy-ion collisions at 200 GeV.
STAR results can provide valuable insights into the interactions between $D^{0}/\overline{D^0}$-$\pi^{\pm}$, $D^{0}/\overline{D^0}$-$K^{\pm}$ and $D^{0}/\overline{D^0}$-$p^{\pm}$ pairs during the hadronic phase. 
$D^0 (\overline{D^0})$ mesons are reconstructed via the $K^{\mp}-{\pi}^{\pm}$ decay channel using topological criteria enabled by the HFT (Heavy Flavor Tracker) detector with excellent track pointing resolution.
These proceedings show a comparison study between STAR results and theory predictions using NLO-HMChPT (Next-to-Leading Order-Heavy Meson Chiral Perturbation Theory) scheme and associated physics implications.   
 }

% Keywords
%\keyword{Heavy-flavour; heavy-ion collision; femtoscopy; Quark-Gluon-Plasma; $D^0$ meson reconstruction; final-state interaction} 

\maketitle
\vspace{-0.5cm}

\section{Introduction and Motivation}
\label{intro}
Ultra-relativistic heavy-ion collisions produce a deconfined state of nuclear matter, known as quark-gluon plasma (QGP), which plays a crucial role in how the system evolves over time. 
Due to a larger mass, heavy quarks, like charm (c) and its charge conjugate ($\overline{c}$) are predominantly produced in the initial hard scatterings, in contrast to the thermal production of lighter quarks (u, d, s) within the hot fireball~\cite{PoS}. 
The presence of charm/anticharm quarks in $D^0$ mesons ($D^{0}$: $c{\overline{u}}$ and $\overline{D^{0}}$: $u{\overline{c}}$) make them excellent probe to study the interaction of charm quarks with the medium by measuring different observables. 
For example, distribution of nuclear modification factor ($R_{AA}$) as a function of transverse momentum ($p_T$) for the $D^0$, $D^+$ and $D^{*+}$ mesons in Au+Au, Pb+Pb and p+Pb systems are studied by both the RHIC and LHC experiments. They observed strong suppression of D-meson production yields at high $p_{\mathrm{T}}$ in large collision systems (Au+Au, Pb+Pb) which indicates presence of hot and dense QGP medium~\cite{Raa, ALICE_RAA}. Small $D$ meson suppression over low $p_T$ region in the p+Pb system is due to cold nuclear matter effects~\cite{ALICE_RAA_p_Pb}.
%needs to update references
%Add ref for p+Pb system
RHIC and LHC data also indicated significant $D^0$ elliptic flow ($v_2$) in heavy-ion collisions~\cite{D0v2, ALICE_RAA, CMS_V2}. 
Several theory predictions with different assumptions were used for comparison but none of those could describe all the data simultaneously. This indicates requirement of new observables to constrain different models which will also help to understand particle production mechanism.
   
To shed more light on this problem, we measured femtoscopic correlation functions between $D^0$ meson and charged hadron ($\pi$, $K$, $p$) pairs using STAR data of Au+Au collisions at $\sqrt{s_{NN}}$ = 200 GeV. 
Femtoscopy is sensitive to the spatial properties of an emitting source, quantum statistics, and interactions in the final state (e.g. Coulomb interactions and final-state strong interactions). The structure of the source is typically expressed in terms of a source function S(r), which represents a time-integrated distribution of emission points. Analytical models, like, Lednicky-Lyuboshitz can relay the experimentally measured correlation function to the emission source size and final state interaction parameters via the scattering amplitude, assuming an effective range approximation~\cite{LL, Lisa}.
Thus, femtoscopic correlations could be used to deduce both the source effective “interaction” volume $V_{eff}$, and the cross section for the interactions in the final state.
Recently, ALICE measured the same observable for charged D-meson and charged hadron ($\pi^{\pm}$, $p^{\pm}$) pairs in $p+p$ collision system~\cite{ALICE_D-pi},~\cite{ALICE_D-p}. These findings encouraged us to learn about final state interactions between neutral D-meson and charged hadron pairs in heavy-ion collisions where QGP appears.
Theory of femtoscopy says, the phase-space cloud of outgoing correlated pairs, also referred to as the area of homogeneity is influenced by the dynamics of QGP, such as collective flow~\cite{Sinyukov, Lisa}. 
In cases of strong correlations, the size of this homogeneity region is significantly smaller than the overall size of the fireball~\cite{Lisa}.

\section{Particle identification and D-meson reconstruction}
STAR consists of several detectors~\cite{STARdetector}, each serving a unique purpose. The Time Projection Chamber (TPC) and the Time of Flight (TOF) detectors are the main components used for tracking and identifying charged particles. For reconstructing $D^0$ and $\overline{D^0}$ mesons through the $K^{\mathrm{\mp}}\pi^{\mathrm{\pm}}$ decay channel, the Heavy Flavor Tracker (HFT) was employed due to its exceptional track-pointing resolution, using a set of topological selection criteria as described in~\cite{Raa}.

\begin{figure}[!ht]
    \centering
    \includegraphics[width=0.45\linewidth]{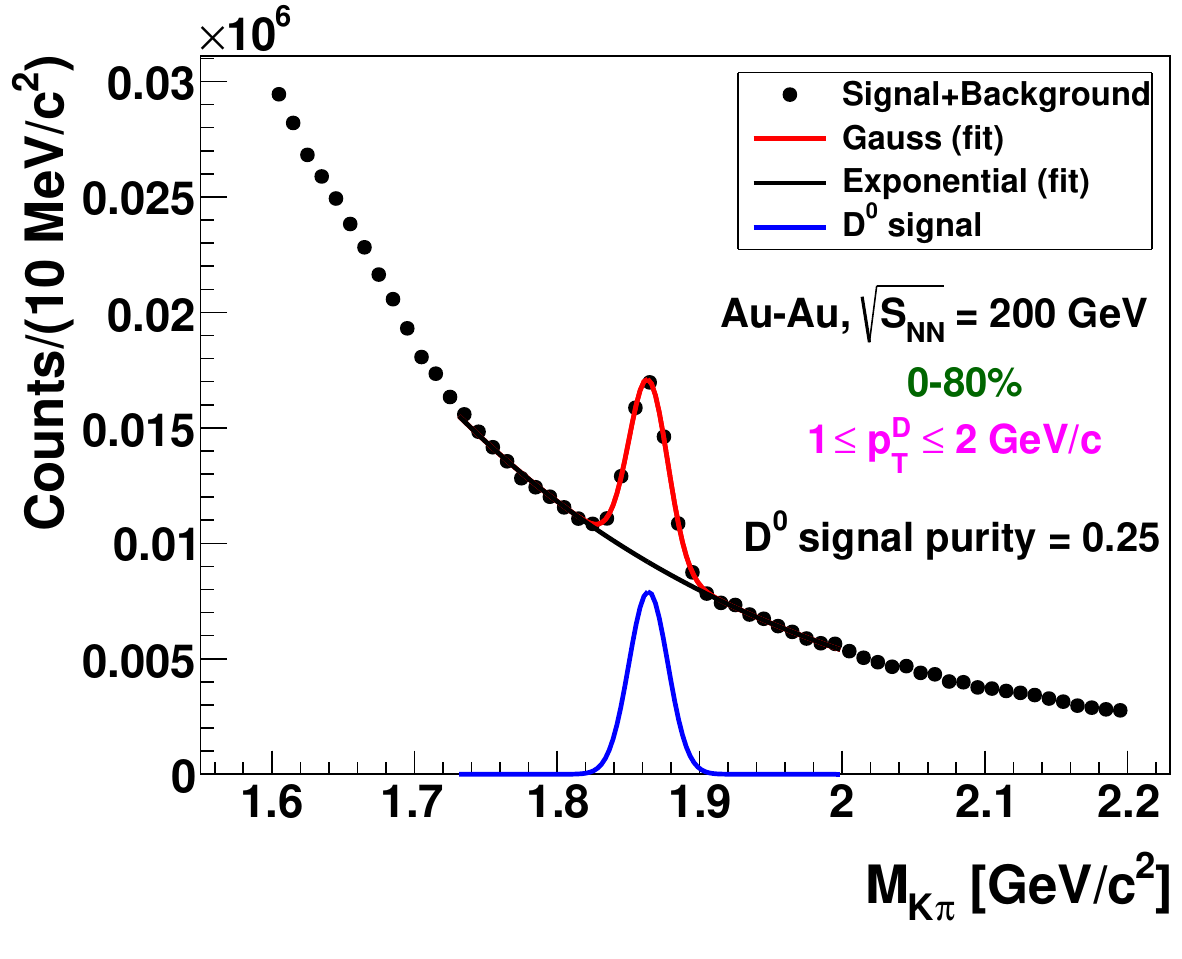}
    \hspace*{0.25cm}
    \includegraphics[width=0.45\linewidth]{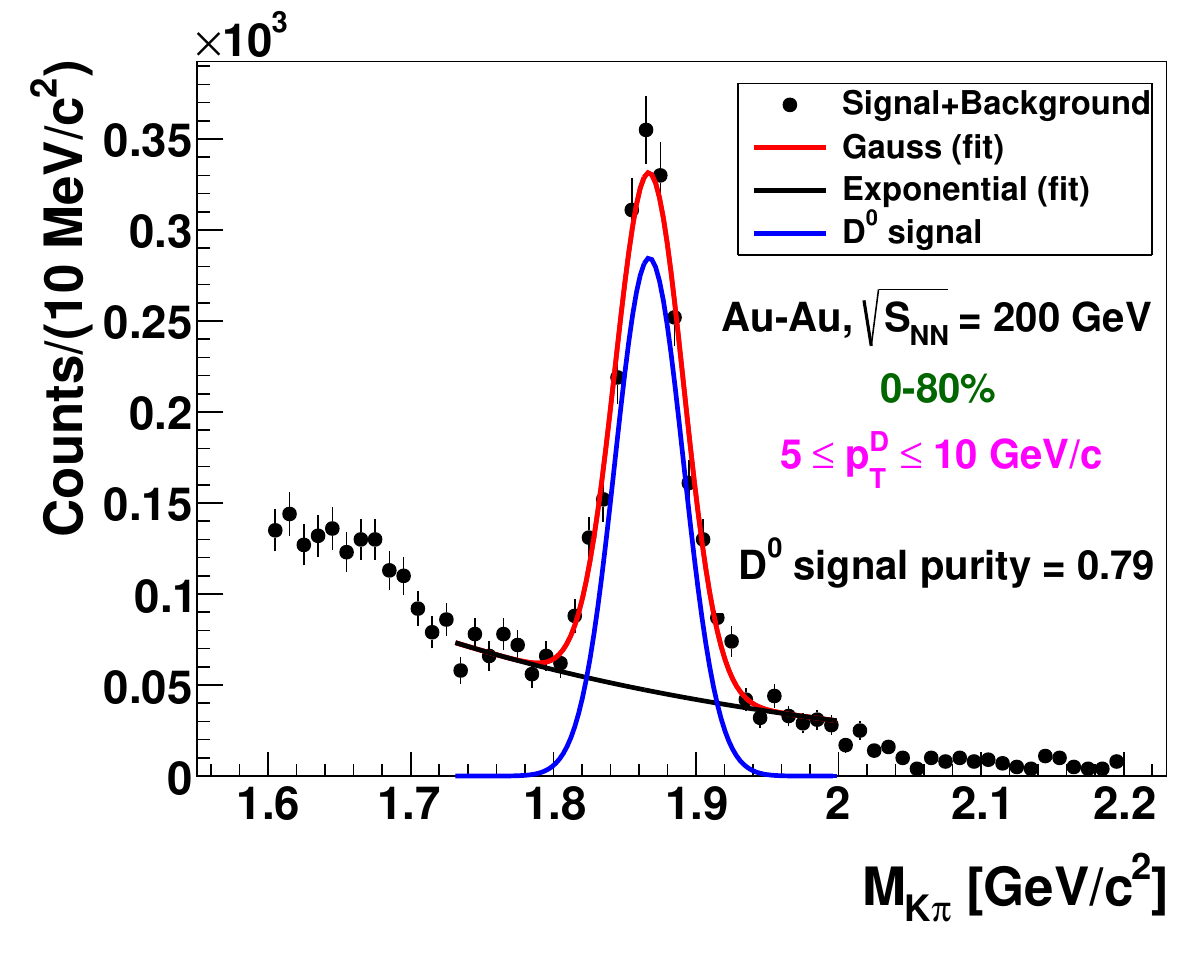}
    \vspace*{0.1cm}
    \caption{The invariant mass ($M_{K\pi}$) distribution of $D^0$ and $\overline{D}^0$ candidates is shown using data from STAR Run 2014. Black solid circles show the mass values for pairs of $K$ and $\pi$ with opposite charges. The red line is a Gaussian fit and the black line shows an exponential fit to the background from these pairs. The blue curve is the fit for the $D^0$ ($\overline{D}^0$) signal between 1.73 and 2.0 GeV/$c^2$.}
 \label{fig:enter-label}
\vspace{-0.8cm}
\end{figure}
\vspace{1.0cm}

Figure 1 presents the invariant mass distributions of $D^0$ and $\overline{D^0}$ candidates in two $p_T$ intervals using STAR data of Au+Au collisions at $\sqrt{s_{NN}}$ = 200 GeV. We achieved high-purity $D^0$ ($\overline{D^0}$) signal within the mass range of 1.82 to 1.91 GeV/$c^2$.
$D^0$ candidates (and their charge conjugates) were chosen based on following criteria: $p_{\mathrm{T}}>$ 1 GeV/$c$ and signal-to-combinatorial background ratio (S/B) exceeding 30\%\ in the lowest $p_{\mathrm{T}}$ bin.
As $p_{\mathrm{T}}$ increases, the S/B ratio also improves. The signal purity, defined as $S/(S + B)$, was evaluated for each $p_{\mathrm{T}}$ bin: 1–2, 2–3, 3–5, and 5–10 GeV/$c$. In the lowest $p_{\mathrm{T}}$ bin, the purity is approximately 25\%\, rising steadily to about 80\%\ in the highest bin. This $D^0$ purity is used to compute the $D^{0}-hadron$ pair purity, as further detailed in Section 3.

\section{Raw and purity-corrected correlation function}
According to theory, the femtoscopic correlation function $C(k^*)$ can be defined by the Koonin-Pratt formalism~\cite{Lisa} as Eq.~\ref{eq:CKP}:

\begin{equation}
C(k^*) = \int S(r)\, |\psi(k^*, r)|^2 \, \mathrm{d}^3r,
\label{eq:CKP}   
%\vspace{-0.1cm}
\end{equation} 
where $S(r)$ is the emission source function and $\psi(k^*, r)$ is the pair wave function. The variable $k^*$ denotes the relative momentum between two correlated particles emitted from a source of spatial extent $r$. In our analysis, $C(k^*)$ was computed as the ratio of the correlated $D^0$-hadron pair distribution, $A(k^*)$, to the uncorrelated pair distribution, $B(k^*)$, in the pair’s center-of-mass rest frame, as described in Eq.~\ref{eq:Cexp}~\cite{Lisa}.

\begin{equation}
 C_{\text{measured}}(k^*) = N\frac{A(k^*)}{B(k^*)} \hspace{2mm} \text{and} \hspace{2mm}  k^* = \frac{1}{2} (p_{\mathrm{1}} - p_{\mathrm{2}}) , 
\label{eq:Cexp}   
%\vspace{-0.1cm}
\end{equation}  
The normalization factor is denoted by $N$, while $p_{\mathrm{1}}$ and $p_{\mathrm{2}}$ correspond to the momenta of the $D^0$ meson and light-hadron tracks, respectively, in the rest frame of the particle pair. To calculate $A(k^*)$, both tracks were selected from the same event. For $B(k^*)$, the event-mixing method was applied, in which tracks from different events having similar primary vertex position ($V_z$) and centrality class, were combined to create uncorrelated pairs.

Accuracy of the measurement of correlation functions can be influenced by the presence of misidentified correlated pairs. To select primary tracks of the charged hadrons under study ($\pi$, $K$, and $p$), both the Time Projection Chamber (TPC) and Time-of-Flight (TOF) detectors were utilized. Possible detector-related effects from hadron-track splitting and self-correlations involving the daughter particles of the $D^0$ were minimized. Although detector artifacts such as the merging of two distinct tracks into one can also occur. Our investigation indicated that the impact of such merged tracks was negligible. Purity corrections were applied to raw correlation functions in order to eliminate the influence of combinatorial background beneath the $D^0$ signal peak, as well the contamination in $\pi$, $K$, and $p$ samples from other hadrons and electrons.  Formula~\cite{PRC74} used for $D^0$-hadron pair-purity corrections:
\begin{equation}
 C(k^*) = \frac{C_{\text{measured}}(k^*) - 1}{\text{Pair Purity}} + 1,
 \label{eq:Ccorr}   
\end{equation}
In this context, $C(k^*)$ denotes the final correlation function after applying the purity correction. $C_{\mathrm{measured}}(k^*)$ refers to the correlation function corrected for potential detector effects. The Pair Purity is obtained by multiplying the $D^0$ signal purity with the average purity of the hadron sample.

The hadron sample purity was evaluated using the standard method employed by the STAR experiment~\cite{nsigma}. Hadrons were selected based on momentum criteria: $p_{\pi} < 1$ GeV/$c$, $p_{K} < 1$ GeV/$c$ and $p_{p} < 1.2$ GeV/$c$, as particle identification becomes increasingly difficult beyond these thresholds due to overlap with electrons and other hadrons. Within the chosen momentum ranges, the average purities for pions and protons are (99.5 $\pm$ 0.5)\%\, while the kaon sample exhibits a purity of (97 $\pm$ 3)\%\ . Systematic uncertainties were assessed by varying the topological selection criteria used in $D^0$ reconstruction~\cite{Raa}, as well as by accounting for uncertainties in the purity estimation of $D^0$-hadron pairs.

\section{Results and Discussion}
This section presents the distribution of final correlation function, $C(k^*)$ followed by pair-purity correction for neutral $D$-meson and identified charged hadron pairs.
Figure 2, from left to right, presents STAR preliminary results on the $C(k^*)$ for $D^0/\overline{D^0}-\pi^{\pm}$, $D^0/\overline{D^0}-p^{\pm}$, and $D^0/\overline{D^0}-K^{\pm}$ pairs respectively. These results are obtained from minimum bias Au+Au collision data at $\sqrt{s_{\mathrm{NN}}} = 200$ GeV within the pseudorapidity range $|\eta| < 1$. The $C(k^*)$ values are largely consistent with unity, suggesting no significant correlation signal, though notable statistical fluctuations are present. The observed correlation strength is directly related to the spatial extent of the particle-emitting source~\cite{Lisa, Sinyukov}.

\begin{figure}[!ht]
    \centering
    \vspace{-0.3cm}
    \includegraphics[width=1.0\linewidth]{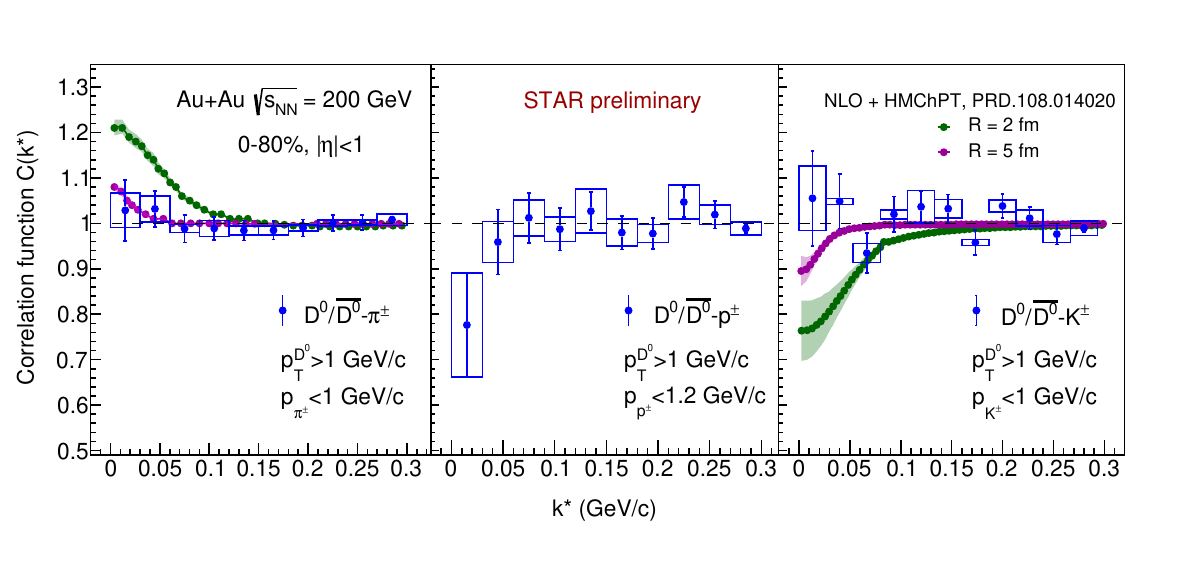}
 \vspace*{-0.8cm}
    \caption{The $C(k^*)$ values are shown for three pairs: (left) $D^0/\overline{D^0}-\pi^{\pm}$, (middle) $D^0/\overline{D^0}-p^{\pm}$, and (right) $D^0/\overline{D^0}-K^{\pm}$. The blue circles with boxes represent the STAR data and the related systematics uncertainties. The green and pink bands show the predicted $C(k^*)$ values from the NLO + HMChPT model for source sizes of 2 fm and 5 fm, respectively~\cite{PRD108}.}
    
 \vspace*{0.3cm}
\end{figure}
The lack or weakness of a correlation implies that the emission source producing the $D$-hadron pairs is relatively large in spatial extent.
Figure 2 also compares STAR data with correlation functions computed using the next-to-leading order (NLO) Heavy Meson Chiral Perturbation Theory (HMChPT) framework~\cite{PRD108, ref1NLO, ref2NLO}. The left panel presents theoretical predictions for $D^0-\pi^+$ and $D^+-\pi^0$ pairs, while the right panel shows results for $D^0-K^+$ pairs~\cite{PRD108}. It is important to note that Coulomb interactions are not included in any of these theoretical channels. 
Comparing to the model predictions, the STAR results are consistent with an emission source radius of 5 fm or larger.
On the left panel, a minimum on the theoretical $C(k^*)$ can be found around 0.215 GeV which refers the presence of lightest $D_{0}^{*}$ state. Although this depletion dilutes with increasing radius of the correlation source~\cite{PRD108}. No bound state was observed in STAR data of $C(k^*)$ for $D^{0}\pi$
pairs.
On the right panel, due to the threshold difference between $D^0 K^+$ and $D^+ K^0$ channel, a cusp effect is observed near 0.083 GeV in the theoretical correlation function for $DK$ channel~\cite{PRD108}. Such effect shows a inversely proportional source size dependence. 
The $C(k^*)$ prediction of the same channel indicates a noticeable suppression in the $k^*$ range from 0 to 0.05 GeV/$c$, attributed to the presence of the $D^*_S(2317)^\pm$ bound state. This suppression becomes more prominent as the emission source radius decreases~\cite{PRD108}. 
However, the STAR data do not clearly show this resonance effect as well as the cusp-like effect, due to either large emission source or sizable statistical uncertainties.

\section{Conclusion}
These proceedings report the first measurement of heavy-flavour femtoscopy in Au+Au collisions by the STAR experiment. The results are consistent with no correlation signal as well as large emission source size. Moving forward, we intend to improve these results by merging data collected during Runs 2014 and 2016. This combined dataset will increase the precision of the correlation function measurements and provide better insights into the source size. Furthermore, incorporating theoretical inputs will be essential for a more comprehensive interpretation of the data.

\vspace{6pt}
\section{Acknowledgement}
Priyanka Roy Chowdhury acknowledges the financial support received from the National Science Centre, Poland (NCN) through grant no. 2018/30/E/ST2/0008, along with partial funding provided by the U.S. Department of Energy (DOE).

\end{document}